\documentclass[a4paper, notoc]{JHEP3}
\usepackage{amsfonts}
\usepackage{amssymb}
\usepackage{epsfig}
\usepackage{cite}

\newcommand{\be}{\begin{equation}}
\newcommand{\ee}{\end{equation}}
\newcommand{\bea}{\begin{eqnarray}}
\newcommand{\eea}{\end{eqnarray}}
\newcommand{\mbb}{\mathbb}
\newcommand{\ti}{\times}
\newcommand{\half}{\frac{1}{2}}
\newcommand{\mc}{\mathcal}

\newcommand{\fn}{\footnote}
\newcommand{\beqa}{\begin{eqnarray}}
\newcommand{\eeqa}{\end{eqnarray}}
\newcommand{\pim}{{\rm Im\,}}

\def\z{\zeta}
\def\vt{\vartheta}

\title{K\"ahler Potentials of Chiral Matter Fields for Calabi-Yau String Compactifications}
\author{Joseph P. Conlon, Daniel Cremades, Fernando Quevedo\\ DAMTP,
  Centre for Mathematical
Sciences,\\
  Wilberforce Road, Cambridge, CB3 0WA, UK.\\

  E-mail:
  \email{j.p.conlon@damtp.cam.ac.uk} ,    
  \email{d.cremades@damtp.cam.ac.uk} ,  \email{f.quevedo@damtp.cam.ac.uk}}
\abstract{The K\"ahler potential is the least understood part of
  effective  $N=1$ supersymmetric theories derived from 
string compactifications. Even at tree-level, the 
   K\"ahler potential  for the physical
  matter fields, as a function of the moduli fields, is unknown 
for  generic Calabi-Yau compactifications and has only
  been computed for simple toroidal orientifolds.
 In this paper we describe how the modular dependence of
  matter metrics may be extracted in a perturbative expansion in the
  K\"ahler moduli. Scaling arguments, locality and knowledge of the
  structure of the physical Yukawa couplings are sufficient to find
  the
relevant K\"ahler potential. 
Using these techniques we compute the `modular
  weights' for bifundamental matter on wrapped D7 branes for
  large-volume IIB Calabi-Yau flux compactifications. We also apply our
  techniques to the 
  case of toroidal compactifications, obtaining results consistent with those present
  in the literature. Our techniques do not provide the complex
  structure
moduli dependence of the K\"ahler potential, but are sufficient to extract
relevant information about the canonically normalised matter fields and the soft
  supersymmetry breaking terms in gravity mediated scenarios.}

\preprint{DAMTP-2006-79 \\ NSF-KITP-06-81\\ hep-th/0609180}

\begin{document}

\tableofcontents

\section{Introduction}

Extracting the form of the effective four-dimensional field theory
corresponding to compactifications of string theory has been one of
the most active areas of research in string phenomenology over the
years
\cite{witten, dkl,  imr, hepth0409098, candelas}. For $N=1$
supersymmetric compactifications\footnote{Although the validity of our
  discussion is general, we will concentrate mostly on  Calabi-Yau
  orientifold compactifications of type IIB string theory.} we know that the effective
supergravity theory depends on the K\"ahler potential $K(\Psi,\Psi^\dagger)$, the
superpotential $W(\Psi)$ and the gauge kinetic function
$f(\Psi)$, where $\Psi$ represents the chiral superfields surviving at
low energies.
These include both the charged matter superfields $C$ and the singlet
moduli superfields $\Phi$. It is well known that $W$ and $f$, being
holomorphic, are under much better control than the real function $K$.
In particular $K$ is not protected by the standard non-renormalization
theorems of $N=1$ supersymmetry.

The standard way to extract the functional form of $K$, $W$ and $f$ at
tree-level is by 
dimensionally reducing the original 10D theory, having carefully
identified the appropriate 4D superfields in terms of the
corresponding 10D geometrical quantities (see for instance
\cite{hepth0409098}). This allows the determination of $K,W$ and $f$
as functions of the moduli fields and some of the matter
fields. However there are some  matter fields for which
dimensional reduction cannot provide the K\"ahler potential. These
include
 the twisted sector fields of orbifold and orientifold
compactifications, bifundamental fields from magnetised D7 branes and those
stretching between D3 and D7 branes. For these cases explicit string amplitudes
need to be computed in order to extract the correct K\"ahler
potential.
These calculations are essentially limited to flat backgrounds such as
toroidal orientifolds \cite{hepth0406092,hepth0412150, lerda}, severely
limiting the information that can be extracted.
In particular an explicit calculation in a IIB Calabi-Yau orientifold seems out of reach.

The importance of knowing the K\"ahler potential for the physical
matter fields is clear: it is needed for correctly identifying the
canonically normalised fields and therefore determines the structure of
most of the observable physical quantities, such as the corresponding
scalar potential, the Yukawa couplings, etc. In particular, the matter
K\"ahler potential plays a crucial role in the determination of soft
supersymmetry breaking terms.

On this regard let us be more specific. 
There has been much recent effort in understanding supersymmetry breaking in string compactifications.
This follows on the progress made in moduli stabilisation
\cite{hepth0105097, hepth0301240}, which allows the moduli potential
to be computed from first principles. The moduli breaking supersymmetry can be identified explicitly and 
their F-terms evaluated. Prior to this and in the absence of explicit
moduli potentials\footnote{Moduli potentials were actually studied in
  the past but without fixing all moduli and with no explicit control
  on hierarchies.}, it was necessary
to parametrise supersymmetry breaking as $S$-, $T$- or $U$- dominated, where
$S$ is the 4D dilaton, $T$ the K\"ahler moduli and $U$ the complex
structure moduli. Scenarios of supersymmetry breaking were then constructed 
and analysed without an explicit moduli potential \cite{bim}. In that the discussion
of supersymmetry breaking now involves explicit moduli potentials, much technical progress has been made.

However one major obstacle in phenomenological analyses has remained, which is the lack of knowledge of the K\"ahler metric
for Standard Model matter fields. 
The computation of soft terms starts by expanding the 
superpotential, metric and gauge kinetic functions
as a power series in the matter fields,
\bea
\label{PowerSeriesExpansion}
W & = & \hat{W}(\Phi) + \mu(\Phi) H_1 H_2 + \frac{1}{6} Y_{\alpha
  \beta \gamma}(\Phi) C^{\alpha} C^{\beta} C^{\gamma} + \ldots, \\
K & = & \hat{K}(\Phi, \bar{\Phi}) + \tilde{K}_{\alpha \bar{\beta}}
(\Phi, \bar{\Phi}) \bar{C}^{\alpha} C^{\bar{\beta}} + \left[ Z(\Phi,
  \bar{\Phi}) H_1 H_2 + h.c. \right] + \ldots, \\
\label{MatterK}
f_a & = & f_a(\Phi).
\eea
$C^\alpha$ denotes a matter field and $\Phi$ a modulus field. In the explicit expressions for soft terms,
the matter metric $\tilde{K}_{\alpha \bar{\beta}}$ plays a crucial role. This quantity is non-holomorphic,
and thus unprotected and hard to compute. However it plays a central role as it 
determines both the normalisation of the matter fields and their
mass basis. In general, an arbitrary form of $\tilde{K}_{\alpha \bar{\beta}}$ can lead to large flavour-changing neutral currents
and off-diagonal A-terms. In the absence of other information, $\tilde{K}_{\alpha \bar{\beta}}$ is often assumed to be diagonal and
moduli-independent. However, this assumption clearly does not hold for
string compactifications where $\tilde{K}_{\alpha \bar{\beta}}$ is
a complicated function of the moduli. Given its
importance for phenomenological applications, obtaining control over the functional form
of $\tilde{K}_{\alpha \bar{\beta}}$ is one of the most important problems in string phenomenology.

As mentioned before, explicit string CFT calculations such as
\cite{hepth0404134} have been used in toroidal compactifications to
work out the matter metrics for adjoint, Wilson line and bifundamental matter.
For Calabi-Yau
cases, dimensional reduction of D-brane actions has allowed the determination of Wilson line and adjoint scalar 
metrics \cite{hepth0409098}. But so far there exist very few results
for K\"ahler metrics for bifundamental matter on Calabi-Yau
backgrounds. These are probably the most important phenomenologically
since these are the standard chiral fields in D-brane models which
will include the quarks and leptons as well as their superpartners.

The purpose of this paper is to give new techniques, applicable to Calabi-Yau backgrounds,
 for computing K\"ahler matter metrics. The approach is to compute the modular dependence of $\tilde{K}_{\alpha \bar{\beta}}$ by
studying the modular dependence of the physical Yukawa couplings. In certain cases this can be determined easily
through dimensional reduction. However in supergravity this is related to the matter metrics, and it is this that will allow
us to determine the modular weights\footnote{We follow standard
 conventions in calling the powers of moduli fields in the K\"ahler
 potential the modular weights of the corresponding matter fields.
 The name came from the transformation
 properties of the corresponding field under (toroidal) 
$T$-duality \cite{cvetic} in heterotic models.} 
 of $\tilde{K}_{\alpha \bar{\beta}}$. Our main application will be to use these
techniques to compute modular weights
for bifundamental matter on wrapped magnetised D7 branes in the 
Calabi-Yau geometries of the large-volume models
of \cite{hepth0502058, hepth0505076}.

This paper is structured as follows. In section \ref{philosophy} we outline the philosophy of our approach.
We describe the calculational approach and the conditions on a modulus for its modular weight to be determined using the 
techniques of this paper. We also present a one-dimensional toy example to illustrate the techniques and show its relation to the
IIB compactifications that are our main interest. In section \ref{comps} we apply our approach first to the large-volume
models of \cite{hepth0502058}. We determine the modular weight of the overall volume and describe how the modular weight
of the small cycles can also be computed under certain assumptions of the brane geometry. 
We then apply the same arguments to the toroidal case. The results we obtain are consistent with the 
explicit computations of \cite{hepth0404134}. In section \ref{conclusion} we conclude.

\section{Philosophy of the Approach}
\label{philosophy}

To simplify the notation we will consider diagonal matter metrics, 
although the argument and results holds for the general case. 
This assumption simplifies (\ref{MatterK}) to
\be
\label{DiagonalK}
K = \hat{K}(\Phi, \bar{\Phi}) + \sum_{\alpha} \tilde{K}_{\alpha}
(\Phi, \bar{\Phi}) C^{\alpha} \bar{C}^{\bar{\alpha}} + \left[ Z(\Phi,
  \bar{\Phi}) H_1 H_2 + h.c. \right] + \ldots.
\ee
Using (\ref{DiagonalK}) we can define the canonically normalised
matter fields $\hat{C}^\alpha$.
These are related to the unnormalised fields $C^\alpha$ by
\be
\hat{C}^\alpha = \tilde{K}_\alpha^\half (\Phi, \bar{\Phi}) C^\alpha.
\ee

The approach we take stems from the supergravity formula for the physical (i.e. normalised)
Yukawa couplings,
\be
\label{PhysicalYukawaCouplings}
\hat{Y}_{\alpha \beta \gamma} = e^{K/2} \frac{Y_{\alpha \beta \gamma}}{(\tilde{K}_{\alpha} \tilde{K}_{\beta} 
\tilde{K}_{\gamma})^{\half}}.
\ee
(\ref{PhysicalYukawaCouplings}) implies that information about the modular dependence of the matter metrics is encoded in the 
modular dependence of the physical Yukawas $\hat{Y}$, which may be relatively easy
to compute directly.
Our aim is to use (\ref{PhysicalYukawaCouplings}) in order to compute the modular dependence of $\tilde{K}_\alpha$.

This approach could yield no useful information if the modular dependence of the superpotential Yukawas
$Y_{\alpha \beta \gamma}$ were unknown. If this were the case, the problem would be overdetermined. We would be unable
to separate the functional dependence of the superpotential Yukawa couplings $Y_{\alpha \beta \gamma}$ and the 
metric dependence $\tilde{K}_\alpha \tilde{K}_\beta \tilde{K}_\gamma$. Even knowing the full functional form of the
physical Yukawa couplings would give no definite information about the functional form of the matter metrics.
However in many cases this dependence is known. Certain moduli are forbidden from appearing in $Y_{\alpha \beta \gamma}$, and in this case
the scaling behaviour of $\hat{Y}_{\alpha \beta \gamma}$ can be 
directly related to that of $\tilde{K}_\alpha$.

Our particular interest here is in IIB flux compactifications. In this case the K\"ahler moduli $T_i$ are forbidden
from appearing in the tree-level superpotential. This can be understood from the Peccei-Quinn shift symmetry
\be
\hbox{Im}(T_i) \to \hbox{Im}(T_i) + \epsilon_i,
\ee
which is unbroken perturbatively. As the superpotential is holomorphic, a perturbative dependence on $\hbox{Re}(T)$ will also
give a perturbative dependence on $\hbox{Im}(T)$, violating the shift symmetry. The non-renormalisation theorems then 
imply that the $T_i$ do not appear in the superpotential - and thus the Yukawa couplings $Y_{\alpha \beta \gamma}$ - to any
order in perturbation theory. It is this that makes it feasible to
compute the modular weights of $\tilde{K}_{\alpha}$ with respect to the 
K\"ahler moduli. The complex structure moduli do however enter the
tree-level superpotential, and so it is not possible to extract any
information about $\tilde{K}_\alpha (U_a)$ from $\hat{Y}_{\alpha \beta \gamma}(U_a)$ (even supposing this could be 
calculated). The techniques of this paper will apply only to those moduli (such as $T_i$) that are forbidden from appearing
in $Y_{\alpha \beta \gamma}$ and not to the moduli (such as $U_a$) that do appear in $Y_{\alpha \beta \gamma}$.

Our main interest is in strings supported on magnetised D7 branes. D7 branes wrap 4-cycles whose size is given by 
$\hbox{Re}(T_i)$. In the dilute flux approximation, the gauge coupling is given by the size of the cycle.
The statement that the gauge theory is weakly coupled is equivalent to the statement that the cycle size is large.
In this case the matter metric can be expanded as a power series in
$\tau_i = \hbox{Re}(T_i)$,
\be
\label{tildeKexp}
\tilde{K}_{\alpha} = \tau_i^\lambda \tilde{K}_0(U_a) + \tau_i^{\lambda - 1} \tilde{K}_1(U_a) + \ldots
\ee
$\tau_i$ contains a factor $e^{-\phi} = g_s^{-1}$ and thus
the higher terms in (\ref{tildeKexp}) can be interpreted as loop corrections. Through
(\ref{PhysicalYukawaCouplings}) the modular weight $\lambda$ determines the
scaling of the physical Yukawa couplings with the cycle volume. Thus
the computation of $\lambda$ reduces to the computation of
the scaling of the physical Yukawa couplings with cycle volume.

The techniques we will use below are:

\begin{enumerate}

\item
In a large volume compactification one of the K\"ahler moduli is
much larger than the other ones and determines the overall volume. We can then
concentrate on the leading power of inverse volume in the K\"ahler
potential. Matter fields are localised on one
of the smaller cycles and so we can use locality to restrict the dependence of
the
K\"ahler potential, as rescaling the volume should not
change the physical Yukawa couplings. As we know the relation between
Yukawa couplings and K\"ahler potentials, this provides information
about the volume dependence of the K\"ahler potentials.

\item
Our fundamental calculational tool is the viewpoint that physical Yukawa couplings arise from the triple overlap of normalised
wavefunctions. Due to the constraints of supersymmetry and holomorphy,
these wavefunctions can only depend in a simple fashion
on the K\"ahler moduli: classically these enter the normalisation only as an overall scale as in (\ref{tildeKexp}). 
The detailed form of the wavefunctions, giving
rise to flavour and textures, come from the complex structure
moduli, which enter into the superpotential $Y_{\alpha \beta \gamma}(U_a)$.
The point is that the overlap integral has a simple dependence on the K\"ahler
moduli and its scaling can be computed without having to compute
  $Y_{\alpha \beta \gamma}(U_a)$.
\end{enumerate}

This understanding of Yukawa couplings as due to the triple overlap of normalised wavefunctions is both intuitive and
supported by explicit calculation. In section \ref{largevol} 
we shall describe below how it arises directly in the dimensional reduction of
higher dimensional Yang-Mills theories - these are the low-energy limits of magnetised brane constructions.

As a warm-up, we illustrate the above approach with a one-dimensional
toy example, pointing out the correspondences between it
and the more realistic IIB Calabi-Yau flux compactifications
subsequently considered.

\subsection{A one-dimensional toy model}

The toy model consists of particle states on a 1-dimensional line $x= -\infty \to \infty$. 
Particles are assumed to be localised about defects on the line located
at $\xi_a$, $\xi_b$ and $\xi_c$. The wavefunctions are assumed to be Gaussian and of equal width $a$.
We take an infinite line for convenience, but because of the rapid
wavefunction falloff we may imagine 
identifying the points $x=-100$ (say) and $x=100$ without affecting the physics.
The normalised wavefunctions are 
\bea
\psi_a(x) & = & \frac{1}{\pi^{1/4} a^{\half}} e^{-\frac{(x - \xi_a)^2}{2a^2}}, \\
\psi_b(x) & = & \frac{1}{\pi^{1/4} a^{\half}} e^{-\frac{(x - \xi_b)^2}{2a^2}}, \\
\psi_c(x) & = & \frac{1}{\pi^{1/4} a^{\half}} e^{-\frac{(x - \xi_c)^2}{2a^2}}.
\eea
The forms of these wavefunctions are illustrated in figure \ref{1dplot1} for
$\xi_a = 1.5, \xi_b = 3, \xi_c =-1.5$ and $a=1$.
\FIGURE{\epsfig{file=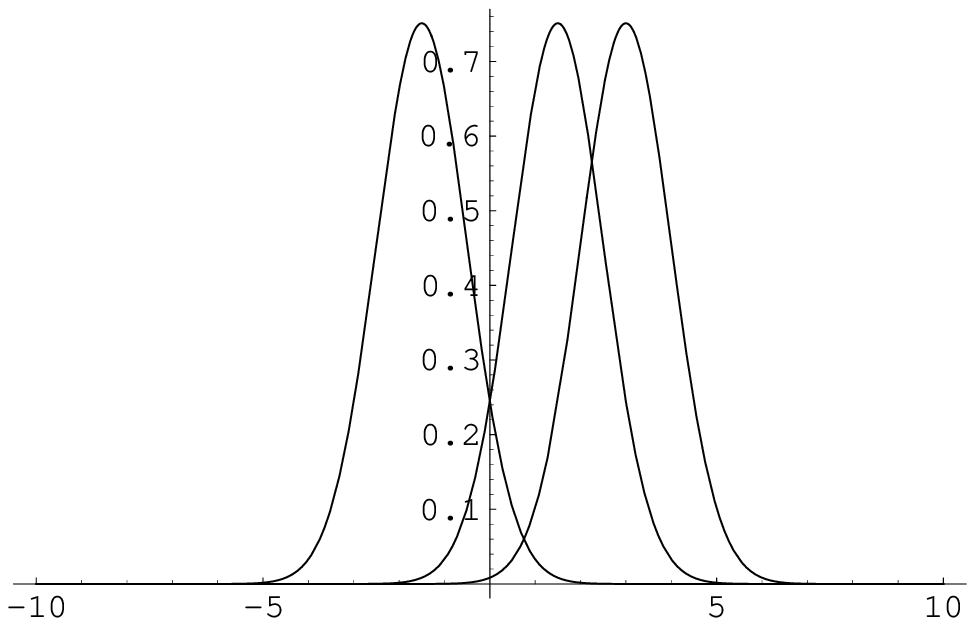}\caption{A one-modulus toy model, illustrating three Gaussian matter wavefunctions. The Yukawa
coupling is determined by the integrated overlap of the normalised wavefunctions.}\label{1dplot1}}
The Yukawa coupling is given by the triple overlap of the normalised wavefunctions,
\bea
\label{1dYuk}
\hat{Y}_{abc} & = & \int_{x=-\infty}^{\infty} dx \, \psi_a(x) \psi_b(x) \psi_c(x) \nonumber \\
& = & \int_{x=-\infty}^{\infty} dx \, \frac{1}{\pi^{3/4} a^{3/2}} e^{\frac{-(x - \xi_a)^2 - (x - \xi_b)^2 - (x - \xi_c)^2}{2a^2}}.
\eea
The Yukawa coupling is exponentially sensitive to the values of the $\xi_i$. In the correspondence with
IIB compactifications, $\xi_i/a$ corresponds to the complex structure
moduli, determining the shape of the wavefunctions, whereas the size of the line corresponds
to the K\"ahler moduli.

We now consider rescaling the size of the line (the `K\"ahler
modulus'), 
without altering the shape of the wavefunctions (the `complex
structure moduli'). This corresponds to scaling
$x \to \lambda x$.
In order for the relative positions and shapes of the wavefunctions to be unaltered, we must also rescale
$\xi_a \to \lambda \xi_a$ and $a \to \lambda a$. 
This leaves the relative breadth and 
central values of the wavefunctions unaltered. 
The new wavefunctions are 
\bea
\psi_a(x) & = & \frac{1}{\pi^{1/4} \lambda^{\half} a^{\half}} e^{-\frac{(x - \lambda \xi_a)^2}{2(\lambda a)^2}}, \\
\psi_b(x) & = & \frac{1}{\pi^{1/4} \lambda^{\half} a^{\half}} e^{-\frac{(x - \lambda \xi_b)^2}{2(\lambda a)^2}}, \\
\psi_c(x) & = & \frac{1}{\pi^{1/4} \lambda^{\half} a^{\half}} e^{-\frac{(x - \lambda \xi_c)^2}{2(\lambda a)^2}}.
\eea
The rescaled wavefunctions (for $\lambda = 2$) are illustrated in
figure \ref{1dplot2}.
\FIGURE{\epsfig{file=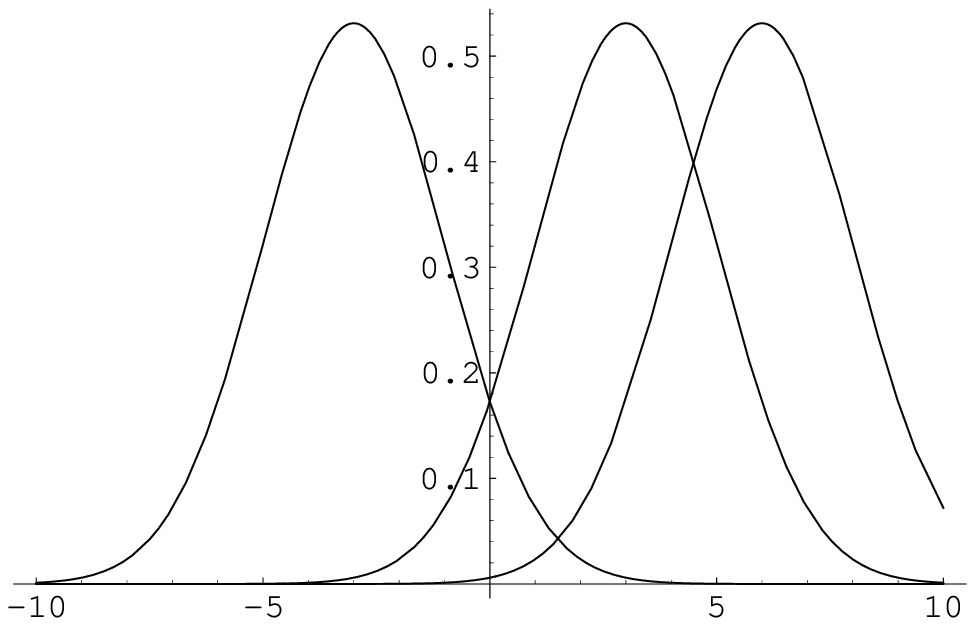}\caption{The same toy model wavefunctions after rescaling the size of the real line. The loci
and breadths of the wavefunctions have also been rescaled and so the
relative shapes are unaltered. 
However the physical Yukawa coupling scales parametrically with the size of the real line and its absolute magnitude 
has changed.}\label{1dplot2}}

However, the physical Yukawa couplings do alter under this rescaling,
\be
\label{1dYukawavariation}
\hat{Y}_{abc} \to \frac{\hat{Y}_{abc}}{\sqrt{\lambda}}.
\ee
Note we can determine the scaling in (\ref{1dYukawavariation}) without ever having to evaluate
the integral in (\ref{1dYuk}). In a IIB context,
computing the integral in (\ref{1dYuk}) corresponds to computing the
complete Yukawa coupling and would require a full-fledged
Calabi-Yau computation. 
However the scaling of the physical Yukawas on the K\"ahler moduli can
be much simpler, and as in (\ref{1dYukawavariation}) we can hope to
compute it through elementary arguments.

In the framework of $\mc{N}=1$ supergravity, 
we could now use the result of (\ref{1dYukawavariation}) to deduce the dependence of the
matter metrics on the `K\"ahler moduli'. However we now seek to move beyond
toy examples and apply the above strategy to IIB Calabi-Yau flux compactifications.

\section{The Large-Volume Model}
\label{comps}

We now apply the above ideas to realistic examples. We will use two
geometries, first that of the large-volume models 
of \cite{hepth0502058, hepth0505076} and then that of the torus. 
As the first involves a full Calabi-Yau geometry, there is
no direct approach to computing bifundamental matter metrics.

\subsection{K\"ahler Metrics}
\label{largevol}

We start this section with a brief description of the geometry of the
large-volume models. 
These models exist within the framework of IIB flux compactifications
with D3 and D7 branes.
The K\"ahler potential and superpotential for the moduli take the
standard form \cite{hepth0301240, hepth9906070, hepth0204254, hepth0403067},
\bea
\hat{K}(\Phi, \bar{\Phi}) & = & - 2 \ln \left( \mc{V} + \frac{\hat{\xi}}{2 g_s^{3/2}} \right) - \ln \left(
i \int \Omega \wedge \bar{\Omega} \right) - \ln (S + \bar{S}). \\
\hat{W}(\Phi) & = & \int G_3 \wedge \Omega + \sum_i A_i e^{-a_i T_i}.
\eea
$\mc{V}$ is the Einstein-frame volume of the Calabi-Yau. 
We use $\Phi$ to denote an
arbitrary modulus field and
do not specify
the total number of moduli. The dilaton and complex structure moduli are stabilised by fluxes. The K\"ahler moduli are
stabilised by a combination of $\alpha'$ corrections and nonperturbative superpotentials. 
As shown in \cite{hepth0502058, hepth0505076}, these very generally interact  
to produce one exponentially large cycle controlling the
overall volume together with $h^{1,1} - 1$ small cycles.
We denote
the
large and small moduli by
$T_b = \tau_b + i c_b$ and $T_i = \tau_i + i c_i$ respectively, with $i = 1 \ldots
h^{1,1} -1$. Consistent with this, we assume the volume can be written as
\be
\mc{V} = \tau_b^{3/2} - h(\tau_i),
\ee
where $h$ is a homogeneous function of the $\tau_{i}$ of degree
3/2. The simplest model, whose properties have been studied in detail
in \cite{hepth0502058, hepth0505076, hepth0605141},
involves the manifold $\mbb{P}^4_{[1,1,1,6,9]}$ and has $h^{1,1} = 2$
with
\be
\mc{V} = \tau_b^{3/2} - \tau_s^{3/2}.
\ee
The large volume lowers both the string
 scale and gravitino mass,
$$
m_s \sim \frac{M_P}{\sqrt{\mc{V}}} \qquad \hbox{ and } \qquad m_{3/2} \sim
m_{soft} \sim \frac{M_P}{\mc{V}}.
$$
The stabilised volume is exponentially sensitive to the stabilised
string coupling, $\mc{V} \sim e^{\frac{c}{g_s}}$, and
may thus take arbitrary values.
A volume $\mc{V} \sim 10^{15} l_s^6 \equiv 10^{15} (2\pi \sqrt{\alpha'})^6$ is required to explain the weak/Planck
hierarchy and give a TeV-scale gravitino mass. As $m_s \gg m_{3/2}$,
the low-energy phenomenology is that of the MSSM and thus the
computation of matter metrics is an important part of the phenomenology.

We will not review the details of the moduli stabilisation here,
leaving those to the references
\cite{hepth0502058, hepth0505076}. Our interest here is rather in computing matter metrics and their dependence on the
geometry. This geometry is illustrated in 
figure \ref{LargeVolumePic}.
\FIGURE{\makebox[15cm]{\epsfxsize=15cm \epsfysize=12cm
\epsfbox{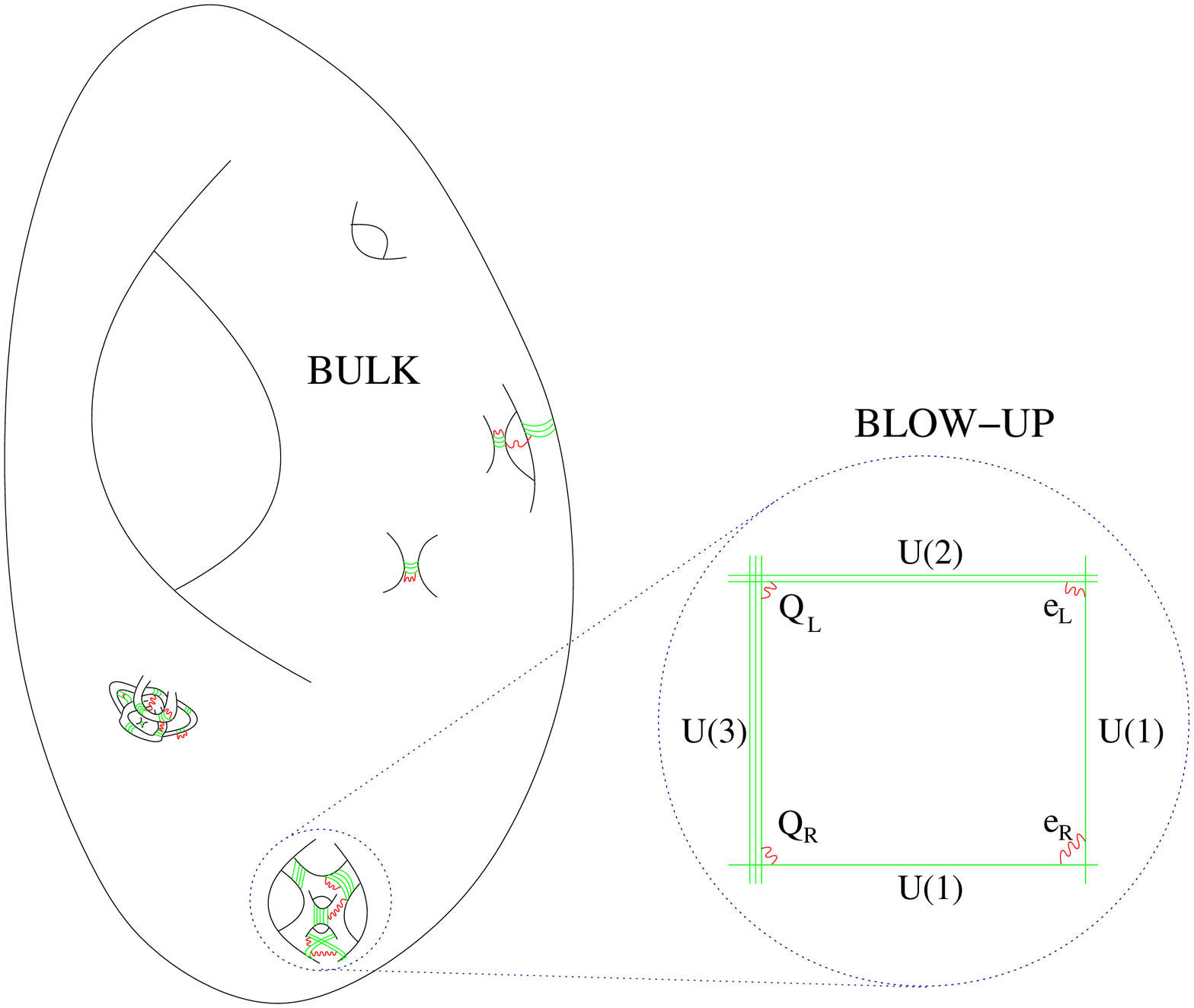}}\caption{The physical picture: Standard Model matter is supported on a
small blow-up cycle located within the bulk of a very large
Calabi-Yau. The volume of the Calabi-Yau sets the gravitino mass and 
is responsible for the weak/Planck hierarchy.}\label{LargeVolumePic}}
The stabilised volumes of the small cycles are $\tau_s \sim \ln \mc{V}$. D7-branes
wrapped on such cycles have gauge
couplings qualitatively similar to those of the Standard Model. 
If the branes are magnetised, Standard
Model chiral matter can arise from strings stretching between stacks
of D7 branes. We assume the Standard Model arises from a stack of magnetised branes
wrapping one (or more) of the small cycles. Given this assumption,
our aim is to compute the modular weights of the matter metrics for the bifundamental chiral matter.

In what we shall call the `minimal model', there exists only one small
blow-up 4-cycle on which a stack of magnetised 
D7 branes are wrapped.
The existence of only one small cycle need not be incompatible with the several
different gauge factors of the Standard Model. The spectrum of chiral
fermions depends on the magnetised flux $F$ present on the brane
worldvolume. This is quantised on 2-cycles $\Sigma_i$,
\be
\int_{\Sigma_i} F \in \mbb{Z}.
\ee 
If several such
2-cycles exist within the 4-cycle, different brane stacks can be realised through
different choices of 2-form flux on these 2-cycles. This is consistent with there being only one small cycle, 
as these 2-cycles may be homologically trivial within the Calabi-Yau and only non-trivial when
restricted to the 4-cycle.
As the cycle is a blow-up cycle, the branes cannot move off the cycle and have no adjoint matter.
This permits a chiral spectrum, as found in explicit models of branes
at singularities \cite{bas}.
The geometry of this minimal model is shown in figure \ref{MinimalModel}.
\FIGURE{\makebox[15cm]{\epsfxsize=15cm \epsfysize=15cm
\epsfbox{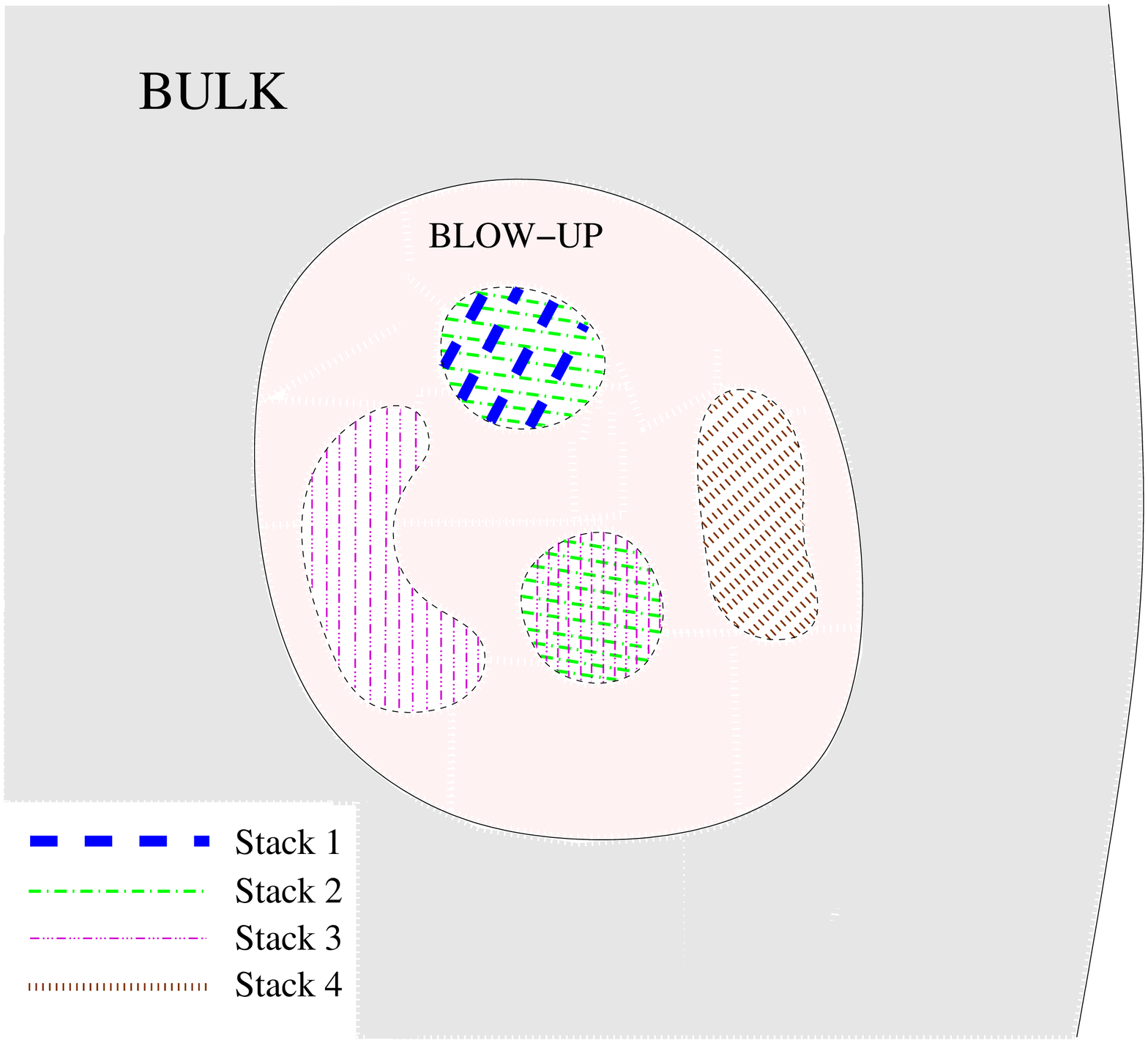}}\caption{
In the minimal geometry, there is only one small 4-cycle. The
  different brane stacks of the Standard Model are distinguished by
  having different magnetic fluxes on the internal 2-cycles of the
  4-cycle. In the minimal model above, these 2-cycles are not inherited from
  the Calabi-Yau and only exist as cycles in the geometry of the
  4-cycle. Four distinct brane stacks are required to realise the
  Standard Model, and we schematically show how these stacks are distinguished by different 
choices of magnetic flux.}\label{MinimalModel}}

We now address the computation of the modular weights.

\subsubsection{Volume Dependence}

In the large-volume limit we can factorise $\tilde{K}_{\alpha}$ as in (\ref{tildeKexp}),
\be
\label{FactorisedK}
\tilde{K}_\alpha = \tau_b^{-p_\alpha} k_\alpha(\tau_i, \phi).
\ee
$\tau_b \sim \mc{V}^{2/3}$ is the size of the large 4-cycle and we use $\phi$ to denote both dilaton and 
complex structure moduli.
While we have included an index $p_\alpha$, from the picture of figure \ref{LargeVolumePic}
we expect universality as locality implies all
matter flavours should see the overall volume in the same way. 
We now argue that actually $p_\alpha = 1$.

To do so let us analyse the expressions for the physical Yukawa couplings. %and $\mu$ term. 
Using $\hat{K} = - 2 \ln \mc{V}$ in (\ref{PhysicalYukawaCouplings}), we obtain
\be
\label{PhysicalYukawas}
\hat{Y}_{\alpha \beta \gamma} = \frac{x Y_{\alpha \beta \gamma}}{(k_{\alpha} k_{\beta} k_{\gamma} (\tau_i, \phi_i))^{\half}}
\tau_b^{\frac{-3 + (p_{\alpha} + p_{\beta} + p_{\gamma})}{2}},
\ee
where $x$ is $\mc{O}(1)$ and defined by $x \mc{V} = \tau_b^{3/2}(1 + \ldots)$.

In the large-volume scenario illustrated in 
figures \ref{LargeVolumePic} and \ref{MinimalModel}, the Standard Model branes are all supported 
around a localised (set of) small cycle(s) within a very large bulk. 
The physical origin of Yukawa couplings is through the 
interaction and overlap of the quantum wavefunctions associated with the different matter fields. 
Matter fields supported on branes are localised on the branes, and thus the wavefunctions for Standard Model matter
all have support in the local geometry on the small 4-cycle. As the interactions are determined only locally,
in the large-volume limit the physical Yukawa couplings should be 
independent of the overall volume, provided that the local geometry is unaltered. 

This argument is equivalent to saying that it is consistent to decouple gravity by taking $M_P/m_s \to \infty$ and
study the field theory on the branes. The decoupling of gravity, which is achieved by taking the volume of the Calabi-Yau
to infinity, does not force the physical Yukawa couplings to vanish. Such a situation is familiar from the study of branes
at singularities, where the low-energy theory on the brane is well-defined and non-trivial even though the Calabi-Yau
is non-compact. 

The effect of the above is to tell us that
the physical Yukawa couplings $\hat{Y}_{\alpha \beta \gamma}$ of (\ref{PhysicalYukawas}) should be invariant under
rescalings $\tau_b \to \lambda \tau_b$. This implies
$$
p_\alpha + p_\beta + p_\gamma = 3
$$
for all matter fields present in the Yukawa couplings. As noted
earlier, to make this deduction it is crucial that the superpotential Yukawa
couplings cannot depend on $T$.
In the scenario of figures \ref{LargeVolumePic} and \ref{MinimalModel}, 
all Standard Model matter arises as localised bifundamental D7-D7 states,
and thus should experience the
overall volume in the same way.
We therefore expect $p_\alpha$ to be universal, giving
\be
\label{p=1}
p_\alpha = 1 \qquad \forall \alpha.
\ee
Equation (\ref{p=1}) completely determines the modular weight of the matter fields with regard to the overall volume.

\subsubsection{Small Cycle Dependence: The Minimal Model}
\label{secYukawas}

We now address the
calculation of the modular dependence on the small
moduli for chiral bifundamental matter. We aim to compute the leading
power-law dependence for the minimal model, working in
the dilute flux approximation.

By performing a series expansion in $\tau_s$, we
can write
\be
\label{MatterExpan}
\tilde{K}_{\alpha \bar{\beta}} = \frac{\tau_s^\lambda}{\mc{V}^{2/3}}
k_{\alpha \bar{\beta}}(\phi).
\ee
As in the dilute flux approximation the $\frac{1}{\tau_s}$ expansion is the perturbative weak coupling expansion, 
we know the expression
(\ref{MatterExpan}) will be valid for large values of
$\tau_s$. Corrections to (\ref{MatterExpan}) subleading in $\tau_s$
will be suppressed at large cycle volume.
The physical Yukawa couplings are given by (\ref{PhysicalYukawaCouplings})
\be
\label{PhyYus}
\hat{Y}_{\alpha \beta \gamma} = e^{\hat{K}/2} \frac{Y_{\alpha \beta \gamma}}
{(\tilde{K}_\alpha \tilde{K}_\beta \tilde{K}_\gamma)^{\half}}.
\ee
To obtain $\lambda$, it is therefore sufficient to obtain the scaling
of $\hat{Y}_{\alpha \beta \gamma}$ with $\tau_s$.

In the minimal model, we assume that the Standard Model comes from dimensional reduction of a stack of D7 branes wrapped 
on the small 4-cycle. 
The chiral spectrum can in principle be found by dimensional
reduction of the higher dimensional super Yang-Mills action in the
presence of magnetic fluxes. This
gives an explicit realisation of the understanding of Yukawas as due to the overlap of normalised
wavefunctions, analogous to the 
computation of Yukawa couplings in either the heterotic string \cite{GSW} or
for D9 branes, for which this problem has been 
treated very explicitly in \cite{hepth0404229}.
The action to be reduced is the DBI action, which in the dilute-flux approximation
reduces to that of super Yang-Mills, whose fermionic terms include
\be
\label{YukInter}
\int_{\mbb{M}_4 \ti \Sigma} \bar{\lambda} \Gamma^i \left( \partial_i + A_i \right) \lambda.
\ee
We drop precise numerical factors of $\pi$ or $\alpha'$.
The higher dimensional gauge field ($A_i$) and gaugino ($\lambda$) can be decomposed in a dimensional reduction,
\be
A_m = \sum_i \phi_{4,i} \otimes \phi_{6,i} \qquad \lambda = \sum_i \psi_{4,i} \otimes \psi_{6,i}.
\ee
We are most interested in the spectrum of massless chiral fermions in four dimensions.
This is determined by counting the number of solutions of the Dirac equation on the cycle in the present of magnetic fluxes.
This is given by an index theorem and is topological, depending only on the cycle geometry and the magnetic fluxes. 
This specifies both the number and charge of the fermions present, and these quantities are invariant under continuous deformations
of the cycle.

Direct reduction of the action (\ref{YukInter}) gives both the kinetic terms
\be
\mc{L}_{kin} \sim \bar{\psi} \partial \psi
\ee
and the Yukawa couplings
\be
\mc{L}_Y \sim \phi \bar{\psi} \psi.
\ee
The magnitude of the physical Yukawa couplings is determined by the relative magnitude of these two terms.
Note that the physical Yukawas are dimensionless quantities and can be determined without any reference to the Planck 
scale or the normalisation of the gravitational action.

A full calculation of the Yukawa couplings requires the explicit scalar and fermion wavefunctions.
We suppose we have solved the Dirac and Laplace equations,
\be
\Gamma^i D_i \psi_A = \Gamma^i D_i \psi_B = \nabla^2 \phi_C = 0,
\ee
where $D_i$ and $\nabla^2$ are the appropriate differential operators on the fluxed 4-cycle. 
From (\ref{YukInter}), the kinetic term for the four-dimensional fermion $\psi_A$ is
\be
 \left( \int_{\Sigma} {\psi}^{\dagger}_{6,A} \psi_{6,A} \right) \int_{\mbb{M}_4} \bar{\psi}_{4,A} \Gamma^\mu \partial_\mu \psi_{4,A}.
\ee
Normalisation of the kinetic terms then requires that
\be
\label{Normalisation}
\int_{\Sigma} \psi_{A,6}^{\dagger} \psi_{A,6} = \int_{\Sigma} \psi_{B,6}^{\dagger} \psi_{B,6} = 
\int_{\Sigma} \phi_6^* \phi_6 = 1.
\ee
The four-dimensional Yukawa couplings are also determined by the action (\ref{YukInter}),
\be
\left( \int_{\Sigma} \bar{\psi}_A \Gamma^i A_{i,C} \psi_B  \right) \int_{\mbb{M}_4} \phi_C \bar{\psi}_A \psi_B.
\ee
The physical magnitude of the Yukawa coupling $\hat{Y}_{ABC}$ are given by the overlap integral
of normalised wavefunctions
\be
\label{Yuss}
\hat{Y}_{ABC} = \int_{\Sigma} \bar{\psi}_A \Gamma^i A_{i,C} \psi_B.
\ee
Our interest is the scaling of (\ref{Yuss}) with the cycle volume.
Under a rescaling $\tau_s \to \beta \tau_s$, it follows from (\ref{Normalisation}) that the normalised wavefunctions
scale as
\be
\label{rescaling}
\psi_A \to \frac{\psi_A}{\sqrt{\beta}}.
\ee
The physical Yukawas then scale as
\be
\label{PhysYukScaling}
\hat{Y'}_{ABC} \sim \int_{\Sigma} (\beta d^4 y) \left( \frac{\psi_A}{\sqrt{\beta}} \right)
 \left( \frac{\psi_B}{\sqrt{\beta}} \right)  \left( \frac{\phi_C}{\sqrt{\beta}} \right)
= \frac{\hat{Y}_{ABC}}{\sqrt{\beta}}.
\ee
One may worry that under rescaling of the cycle volume the wavefunctions would undergo far more dramatic changes
than the simple rescaling of (\ref{rescaling}). In the limit of dilute fluxes and large cycle volumes, 
this cannot occur. If the wavefunctions were to change their shape, rather than just their normalisation, the 
physical Yukawa couplings would also change far more dramatically than the simple scaling of (\ref{PhysYukScaling}).
However, this cannot occur. The texture of the Yukawa couplings comes from the superpotential, and thus
cannot depend on the K\"ahler moduli. They can only be changed by a change in the complex structure moduli, which
has not occurred. The K\"ahler moduli can only affect the physical Yukawa couplings through the power $\lambda$ of
(\ref{MatterExpan}), which corresponds purely to an overall scaling of
 the wavefunctions and not to a change in the shape.

As the cycle size becomes smaller, quantum corrections due to the gauge group on the brane become important. These can
alter the shape of the various wavefunctions - this corresponds to subleading powers of $\tau_s$ in (\ref{MatterExpan}).
However, in the limit of large cycle volumes and dilute fluxes, this effect goes away and the wavefunctions become the purely 
classical ones with scaling behaviour given by (\ref{rescaling}).

The result (\ref{PhysYukScaling}) implies that the scaling of the physical Yukawas with the cycle volume is
\be
\hat{Y}_{\alpha \beta \gamma} \sim \frac{\hat{Y}_{\alpha \beta \gamma}}{\sqrt{\tau_s}}.
\ee
This same dimensional reduction implies that the physical Yukawas do not scale with the overall volume.
This is a calculational illustration of our earlier point that the Yukawa interaction is local and 
so should be insensitive to the bulk volume.

Comparison with equation (\ref{PhyYus}) then shows that the matter metric must scale as
\be
\label{diagscale}
\tilde{K}_{\alpha}(\tau_s) \sim \frac{\tau_s^{1/3}}{\mc{V}^{2/3}} k_\alpha(\phi).
\ee
Here we note that nothing in our analysis has depended on whether the
matter metric is diagonal or otherwise. The flavour structure is encoded in
the superpotential and thus is only seen by the complex structure moduli. We can perform a rotation of the matter fields
to diagonalise the kinetic terms, absorbing the non-diagonality in the
(unknown) Yukawa couplings. Thus the scaling behaviour of
(\ref{diagscale}) also applies to general non-diagonal metrics.
For the minimal model, this therefore determines the matter metric in the large cycle volume dilute flux approximation to be of the form
\be
\label{FullMattMetric}
\tilde{K}_{\alpha \bar{\beta}} = \frac{\tau_s^{1/3}}{\mc{V}^{2/3}} k_{\alpha \bar{\beta}}(\phi).
\ee
While the powers in (\ref{FullMattMetric}) are in principle only the leading terms in a 
power series expansion, they dominate in a 
weak coupling expansion.

How large are the subleading terms? As in the dilute flux approximation $\tau_s$ controls the gauge coupling on the branes,
we should interpret the series expansion in $\tau_s$ as an expansion in the coupling of the gauge theory. For a theory
with gauge coupling 
$$
\alpha = \frac{g^2}{4 \pi},
$$
loop effects are suppressed by a factor $\frac{\alpha}{2 \pi} \equiv \frac{g^2}{8 \pi^2}$. For wrapped D7 branes, reduction of the
DBI action gives 
\be
\alpha = \frac{1}{2 \tau},
\ee
and so loop corrections are suppressed by $\sim 4 \pi \tau \sim 100$ and are at the percent level. Thus this suggests that the 
expansion in powers of $\tau$ is a well-controlled expansion. As the size of $\tau$ is determined by matching onto 
the observed gauge couplings, this is simply the statement that the 
Standard Model gauge couplings at $\Lambda \sim 10^{11} \hbox{GeV}$
(in the large-volume models, this is the string scale required for
TeV-scale supersymmetry) lie deep in the perturbative regime.

Let us discuss the assumptions made in deriving (\ref{FullMattMetric}). The first assumption was that of locality -
the strength of the physical Yukawa interaction is insensitive to the overall volume. The justification for this
is that all chiral matter is localised around the small cycle and thus the interactions are localised as well. 
This assumption completely determines the power of the volume that appears in (\ref{FullMattMetric}). The second assumption
was that of the minimal model - all chiral matter arises from dimensional reduction of a single stack of magnetised branes. 
This determined the power
$\tau_s^{1/3}$ in (\ref{FullMattMetric}). 
It seems difficult to escape the first assumption. However if the local geometry is more complicated than 
that of the minimal model, this second assumption may not hold. 
We now investigate some other possibilities for the local geometry and how they would alter the power $\lambda$.

\subsubsection{Small Cycle Dependence: More complicated geometries}

In the previous section, we assumed all D7s giving rise to the Standard Model wrapped an identical 4-cycle.
If this does not hold, we would expect that (\ref{FullMattMetric}) could be altered. 
We can envisage a situation in which 
the D7s are wrapping different small 4-cycles that are however localised in a region of the CY, with
volumes that are small and approximately equal.
Under this assumption, one can still obtain approximate but concrete 
expressions similar to (\ref{FullMattMetric}).

While we cannot now just reduce a single higher-dimensional action, we again expect that Yukawa couplings will arise
from the overlap of normalised wavefunctions. These wavefunctions are supported on the pair-wise intersection locus of  
D7 branes, while the Yukawa coupling is supported on the triple intersection locus.

\subsubsection*{Three D7 branes intersecting at a point}

Since one would not expect Yukawa couplings to arise in IIB string theory from non-intersecting 
branes\fn{Contrary to the IIA case, in which Yukawa couplings can be generated among three D6-branes with no common intersection by world-sheet instanton amplitudes \cite{hepth0011132}, these kind of contributions cannot appear in IIB orientifolds. The reason is that any world-sheet instanton contribution to the superpotential must be holomorphic in $\int_\Sigma (J+iB)$, $\Sigma$ being the relevant (area minimising) 2-cycle wrapped by the world-sheet, $J$ the Kahler form and $B$ the $B$-field. But in IIB orientifold constructions the internal $B$-field is projected out and hence these contributions are absent. } the minimal scenario is three 
stacks of D7s (named for concreteness $a$, $b$ and $c$), each wrapping a small 4-cycle in the Calabi-Yau, intersecting 
pairwise in 2-cycles (labeled $ab$, $bc$ and $ca$) and whose triple intersection is a single point.  
The wavefunctions corresponding to the chiral fermions arising in the overlap of each pair of stacks have support 
only in the intersection 2-cycle, and hence their dependence  on the 2-cycle volume must be
\beqa
\psi_{ij}^\alpha(z_{ij})\sim{1\over \sqrt{A_{ij}}},
\eeqa
with $z_{ij}$ the complex coordinate parametrising the 2-cycle, $ij=ab,bc,ca$, and $A_{ij}$ is the 2-cycle volume. $\alpha$ 
labels the family replication of the corresponding wave functions.
The interactions of these are distinguished, as already emphasised, only by the complex 
structure moduli. Assuming that the
triple intersection point is given by $(z_{ab}^0,z_{bc}^0,z_{ca}^0)$, the Yukawa coupling will just be the product of the 
wave functions evaluated at this point:
\beqa
\hat Y_{\alpha\beta\gamma}=\psi_{ab}^\alpha(z_{ab}^0)\psi_{bc}^\beta(z_{bc}^0)\psi_{ca}^\gamma(z_{ca}^0)\sim 
{1\over\sqrt{A_{ab}A_{bc}A_{ca}}}\sim \tau_s^{-3/4},
\eeqa
where we have further assumed that all volumes are of similar size and are related to some 
characteristic 4-cycle volume $\tau_s$. Assuming the same behaviour as
in (\ref{MatterExpan}), we readily find $\lambda=1/2$ and hence
\beqa
\label{BI}
\tilde K_\alpha \sim {{\tau_s}^{1/2}\over {\cal{V}}^{2/3}} k(\phi).
\eeqa
The power of $\lambda$ is increased compared to the case of the minimal model.
As the branes wrap different cycles in this example, we would expect that `$\tau_s$' as appears in (\ref{BI}) should be expanded to be a
function of the several moduli corresponding to the different cycles. 

%Incidentally, this is the kind of behaviour one finds in the toroidal case, where three 4-cycles generically 
%intersect at a point, cf. \cite{hepth0404134,hepth0412150}. Notice however that in a torus it is not possible 
%to wrap three space-filling D7-branes in a small, localised area of the whole compactification manifold  and, 
%at the same time, having them intersecting at a point. The torus hence does not belong to the kind of large 
%volume compactifications studied in this paper and there is no reason to expect this  correct dependence on 
%the K\"ahler moduli to be found by these methods . We will tangentially analyse the toroidal case in section ???.
%
%Notice also that the case of the D7s intersecting at a finite number of points can be analysed in an analogous way and yields a similar result.

Another possibility is to have three stacks of branes whose common intersection is a 2-cycle.
There are several possibilities here, some of them not easy to analyse, but there are two cases whose 
behaviour can be extracted straightforwardly. We follow the notation of the previous subsection.

\subsubsection*{Two branes overlapping on a 4-cycle}
Suppose branes $a$ and $b$ overlap on a 4-cycle $\Pi_{s}$ whose volume is given by the K\"ahler modulus $\tau_s$, 
and brane $c$ wraps a different 4-cycle $\Pi_a$ whose intersection with $\Pi_s$ is a 2-cycle whose area is 
denoted by $A$. The corresponding wave functions scale as
\beqa
\psi_{ab}&\sim& {1\over \sqrt{\tau_s}},\nonumber\\
\psi_{bc}&\sim& {1\over \sqrt{A}},\\
\psi_{ca}&\sim& {1\over \sqrt{A}}.\nonumber
\eeqa
Hence the Yukawa coupling scales as
\beqa
\hat Y_{\alpha\beta\gamma} = \int_{\Pi_A} \psi_{ab}\psi_{bc}\psi_{ca} \sim {1\over \sqrt{\tau_s}}.
\eeqa
From this result we again get $\lambda=1/3$ and the dependence of the K\"ahler metric as
\beqa
\tilde K_\alpha \sim {{\tau_s}^{1/3}\over {\cal{V}}^{2/3}} k(\phi).
\eeqa

\subsubsection*{Three branes intersecting pairwise on the same 2-cycle}

We now suppose we have three branes wrapping different cycles, such that the any pair of these branes intersect
in the same 2-cycle $\Sigma$. The three stacks therefore also intersect in $\Sigma$. 
Clearly the three wave functions scale as
\beqa
\psi_{ij}\sim{1\over \sqrt{A}},
\eeqa
where $A$ is the area of $\Sigma$. We find
\beqa
\hat Y_{\alpha\beta\gamma} = \int_{\Sigma} \psi_{ab}\psi_{bc}\psi_{ca} \sim {1\over \sqrt{A}}\sim{\tau_s^{-1/4}}.
\eeqa
Here $\tau_s$ is the four-dimensional volume of a characteristic local 4-cycle of the construction, 
such that (roughly) $A\sim \sqrt{\tau_s}$. We obtain $\lambda=1/6$ and
\beqa
\tilde K_\alpha \sim {{\tau_s}^{1/6}\over {\cal{V}}^{2/3}} k(\phi).
\eeqa
As above we expect that due to the several cycles wrapped $\tau_s$ should be expanded to be a function of the
several moduli corresponding to the different cycles.

\subsubsection*{A bound on $\lambda$}
In all the constructions analysed above we have found a value for $\lambda$ between 0 and 1. 
One could ask whether there is a physical reason for having $\lambda$ within these limits. In fact this does seem to 
be the case.

From the above analyses the physical Yukawa couplings scale as
\beqa
\label{scaling_yuk}
\hat Y \sim \frac{V_{123}}{\sqrt{V_{12}V_{23}V_{31}}}.
\eeqa
$V_{123}$ is the volume of the brane triple intersection, and $V_{ij}$ the volume of the pairwise intersection
between stack $i$ and $j$. 
The $V_{ij(k)}$ are non-decreasing functions of the characteristic small 4-cycle volume $\tau_s$ 
(parametrised at first order by powers $V_{ij(k)} \sim\tau^\alpha$, with $\alpha=1$ if the relevant 
intersections are 4-cycles, $\alpha=1/2$ if they are 2-cycles, etc). Note that, for a given value 
of $\tau_s$, $V_{ijk} \subset V_{ij}$, since $V_{123}$ characterises the volume of the triple intersection. 
Then, if we parametrise the scaling of the Yukawa coupling as 
\beqa
\hat Y \sim \tau^{-\beta}
\eeqa
for some real $\beta$, we see that necessarily $\beta \ge 0$. Now, assuming a dependence of the K\"ahler metrics 
of the fields with $\tau_s$ like $\tilde K \sim \tau_s^\lambda$, we find $\lambda=2\beta/3$, and hence $\lambda\geq 0$.

We can also extract an upper bound on $\lambda$ by similar arguments. An upper bound on $\lambda$ implies an 
upper bound on $\beta$. This will be attained whenever the numerator in (\ref{scaling_yuk})
is minimised and the denominator maximised. Clearly the 
denominator is maximised whenever all $V_{ij} \sim \tau_s$, 
and the numerator will be minimised when $V_{123} \sim 1$. Irrespective of whether this can be realised or not, 
this is clearly the strongest dependence 
possible, since $V_{123}$ cannot scale negatively with the volume. This dependence implies $\beta \leq 3/2$ 
and $\lambda\leq 1$. Hence we conclude that $\lambda \in
[0,1]$.\footnote{The bound $\lambda < 1$ also follows from the
  requirement of a good classical limit, $\tau_b \to \infty, \tau_s
  \to \infty, \tau_b/\tau_s \hbox{ constant}$, in which the K\"ahler
  metric does not diverge.}

\subsection{Vanishing of the $\mu$ term}

We now also argue, using similar scaling arguments as above,
 that in the large-volume models the superpotential $\mu$ term should vanish. Going from supergravity to field theory,
the physical (i.e. normalised)  $\mu$
parameter is given by
\be
\label{PhysicalMuTerm}
\hat{\mu}  =  \left( e^{\hat{K}/2} \mu + m_{3/2} Z - F^{\bar{m}} \partial_{\bar{m}} Z \right)
(\tilde{K}_{H_1} \tilde{K}_{H_2})^{-\half},
\ee
where the $F$-terms are given by:
 \be
F^m = e^{\hat{K}/2} \hat{K}^{m \bar{n}} D_{\bar{n}} \bar{\hat{W}}.
\ee

We write 
\bea 
\tilde{K}_{H_1} & = & \tau_b^{-p_1} k_{H_1}(\tau_i), \\
\tilde{K}_{H_2} & = & \tau_b^{-p_2} k_{H_1}(\tau_i), \\
Z & = & \tau_b^{-p_z} z(\tau_i).
\eea
We do not yet impose $p=1$ because this is helpful in seeing the calculational structure.
The physical $\mu$ term is then from (\ref{PhysicalMuTerm}) found to be
\bea
\hat{\mu} & = & \left( e^{K/2}\mu + m_{3/2} Z - F^{\bar{m}}
\partial_{\bar{m}} Z \right) \left( \tilde{K}_{H_1} \tilde{K}_{H_2}
\right)^{-\half} \\
& = & \frac{\tau_b^{\frac{p_1 + p_2}{2}}}{(k_{H_1} k_{H_2} (\tau_i))^{\half}}
\left( e^{\hat{K}/2} \mu + m_{3/2} Z - F^{\bar{m}} \partial_{\bar{m}}
Z \right) \\
& = & \frac{x \tau_b^{\frac{p_1 + p_2 -3}{2}}}{(k_{H_1}
  k_{H_2})^{\half}} \mu + \frac{z}{(k_{H_1} k_{H_2})^\half}
\tau_b^{\frac{p_1 + p_2}{2} - p_z} m_{3/2} - (F^{\bar{m}} \partial_{\bar{m}} Z) \frac{\tau_b^{\frac{p_1 + p_2}{2}}}{(k_{H_1}
k_{H_2})^{\half}}.
\eea
We evaluate
\be
F^{\bar{m}} \partial_{\bar{m}} Z = p_z m_{3/2} \tau_b^{-p_z} z + \tau_b^{-p_z} F^{\bar{i}} \partial_{\bar{i}} z.
\ee
Therefore
\be
\label{PhysicalMu}
\hat{\mu} = x \frac{\tau_b^{\frac{p_1 + p_2 -3}{2}}}{(k_{H_1} k_{H_2})^{\half}} \mu +
\frac{\tau_b^{\frac{p_1 + p_2}{2} - p_z}}{(k_{H_1} k_{H_2})^{\half}} \left[ z(1-p_z) m_{3/2} - F^{\bar{i}} \partial_i z \right].
\ee

(\ref{p=1}) now implies that the superpotential $\mu$ term must vanish. By using (\ref{p=1}) in 
(\ref{PhysicalMu}), we see that the
volume scaling of the first term of (\ref{PhysicalMu}) is 
$$
\mu' \sim \mc{V}^{-1/3} \mu + \ldots.
$$
However, recall that the string scale behaves with 
volume as 
$$
m_s \sim \mc{V}^{-1/2} M_P.
$$
Thus for any non-zero value of $\mu$
we can make the physical mass $\mu'$ arbitrarily greater than the string scale by taking the classical large-volume
limit $\mc{V} \to \infty$. As such behaviour is unphysical, the only consistent case is 
$\mu = 0$.
Of course, the vanishing of the superpotential
$\mu$ term in (\ref{PhysicalMu}) does not imply the vanishing of the physical $\mu$ term, which can be generated by a non-zero $Z$ in the
Giudice-Masiero mechanism\cite{GiudiceMasiero}.

\section{The Single K\"ahler Modulus Case}

In this section we restrict to the simplest case of one K\"ahler modulus. 
In particular, this is the original realisation
of the KKLT scenario. We consider this case separately for two
reasons. First, as the simplest case it is more often used in the
literature and therefore it is useful to have an explicit expression for
the K\"ahler metric for it. Secondly, the large volume scenario
usually requires more than one K\"ahler modulus and therefore the
results of the previous section do not directly apply to this case.
In particular we cannot just assume the configuration of figure 3.

We can see here that the arguments of section (\ref{secYukawas})
can still be used for the minimal model in which the Standard Model comes
from dimensional reduction of a stack of D7 branes wrapped on a
single 4-cycle of size $\tau=\frac{T+T^*}{2}$. 
The K\"ahler potential then can be
written as
\be
K\ =\ -3\log\left (T+T^*\right) + \tilde{K}(T,T^*)\, C^* C +
\cdots,
\ee
with $\tilde{K}(T,T^*) =\left(T+T^*\right)^{-p}$. 
 We are left with the task of determining the power $p$.

Following section (\ref{secYukawas}) we can see again that the
physical Yukawa couplings scale like
$\left(T+T^*\right)^{-1/2}$
from overlapping wavefunctions. Then, using (\ref{PhyYus}),
$e^{{\hat K}/2}= \left(T+T^*\right)^{-3/2}$ and the fact that the
original superpotential Yukawa couplings $Y$ do not depend on $T$, we get $p=2/3$.

Therefore the K\"ahler potential to leading order in the K\"ahler
modulus expansion looks like:
\be
\label{onemod}
K\ =\ -3\log\left (T+T^*\right) +  \frac{C^* C}{ \left(T+T^*\right)^{2/3}}\  +
\cdots
\ee

Notice that this argument did not use the exponentially large volume.
Furthermore, it can  easily be seen that this power $2/3$ will also
appear in the large volume scenario if the D7 branes wrap the
exponentially large cycle instead of a `small' cycle as was assumed
in the previous section. The reason for this is that the
volume is dominated by the large modulus $\tau_b$ with ${\cal{V}}\sim
\tau_b^{3/2}$, and therefore the K\"ahler potential for the multi-moduli
case looks similar to the one in (\ref{onemod}). 
This is also  consistent with substituting
$\tau_s$ by $\tau_b$, with 
$\lambda=1/3$
and ${\cal{V}}\sim \tau_b^{3/2}$ in (\ref{diagscale}).

\section{Toroidal Examples}
\label{tori}

In this section we apply a similar approach to the case of toroidal compactifications.
This differs from the large volume setup considered earlier, since here it is not possible
to localise a small 4-cycle within a large bulk.
However, we will see how 
one can still get the correct dependence on the K\"ahler moduli for the matter K\"ahler metrics 
from the type of scaling arguments used earlier. Our results can 
be compared with the explicit, complete expressions obtained in \cite{hepth0404134,hepth0412150}. 

Consider a factorisable $T^6$ with three K\"ahler moduli denoted by $t_i$. These are related
to the areas of the 2-tori by $t_i \sim A_j A_k$. Consider a system of three magnetised D7 branes, 
each wrapping a different pair of tori and being point-like in third one\fn{In standard notation, 
we call these a branes $D7_1$, $D7_2$, $D7_3$, where $D7_1$ is point-like in the first torus and wraps the second and third tori.}. 
Being magnetised branes, chiral 
fermions arise from the overlap of each pair of branes. 
For example in the case of $D7_1$ and $D7_2$ branes, this fermion has support on the third torus.
The corresponding normalised wave functions (only defined 
on the overlap between the branes) are given by\fn{In the notation of \cite{hepth0404229} these wave 
functions would have been defined multiplied by two `square roots of $\delta-$functions'. These $\sqrt{\delta}$ 
functions allow the Yukawa couplings to be defined as an integral over the whole $T^6$, rather than only over the 
overlap space. We prefer to remove these delta functions for clarity and for notational consistency with the rest of the paper.}
\beqa
\psi_{12}(z_3)&\sim&{1\over \sqrt{A_3}},\nonumber\\
\psi_{23}(z_1)&\sim&{1\over \sqrt{A_1}},\\
\psi_{31}(z_2)&\sim&{1\over \sqrt{A_2}},\nonumber
\eeqa
where the $z_i$ are the corresponding complex coordinate of each
torus. For clarity, we have dropped the wavefunction dependence on complex
structure moduli which differ between flavours, but actually this can
be explicitly computed. 
This calculation was performed in \cite{hepth0404229}; the generic, complete form of the wave functions is given by
\be
\psi_{ij}^\alpha(z) =  \left({2 \pim \tau |M|\over A_k^2} \right)^{1/4} e^{i \pi  M (z + \z) {\pim (z + \z) \over \pim \tau}} \cdot
\vt
\left[
\begin{array}{c}
\frac{\alpha}{M} \\ 0
\end{array}
\right]
\left( M (z + \z),  M \tau \right).
\label{psisoln}
\ee
In this expression, $z$ stands for the complex coordinate in the
$k^{th}$ torus, 
$\tau$ is the complex structure of this $k^{th}$ torus and $\z$ are
complexified Wilson lines 
degrees of freedom, depending also on the complex structure. $M\in
{\bf Z}$ is the relative magnetic 
flux in the $k^{th}$ torus, and $\alpha$ labels the different matter
fields in the same family. 
$\vt$ is given by the Jacobi theta-function
\be
\vt \left[
\begin{array}{c}
a \\ b
\end{array}
\right] (\nu,\tau) =  \sum_{l \in {\bf Z}} 
e^{\pi i (a + l)^2 \tau} \ e^{2\pi i (a + l)(\nu + b)}.
\label{theta}
\ee
These wave functions are solutions both of the Dirac and Laplace
equations on the magnetised torus.
The purpose of including the wavefunction (\ref{psisoln}) is to
emphasise the contrast between the functional dependence on the K\"ahler moduli $A_k$ and
the complex structure moduli $\tau$.

If we suppose, without loss of 
generality, that the intersection point between the three D7s is located at $z_i=0$, then the physical Yukawa coupling is
\beqa
\label{yukis}
\hat Y_{\alpha\beta\gamma} \sim \psi_{12}(0)\psi_{23}(0)\psi_{31}(0)\sim{1\over \sqrt{A_1A_2A_3}}\sim{1\over (t_1t_2t_3)^{1/4}}.
\eeqa

We see that the product of the three wave functions is always proportional to 
$(t_1 t_2 t_3)^{-1/4}$, and (comparing with the explicit wave functions on (\ref{psisoln})), 
this is the only place where the Kahler moduli appear, as expected. 
Hence, whereas the Kahler moduli only give rise to an overall scale of masses, the complex 
structure moduli are responsible for the structure of eigenvalues that eventually gives rise to the flavour structure of the model.
 We must emphasise 
that this is the first term in a volume expansion for the $A_i$ and subleading contributions are to be expected,
corresponding to quantum corrections to the wave function. 

Let us try and derive the K\"ahler moduli dependence of the matter metrics. The K\"ahler potential for a torus is
\be
K = - \log(s + \bar{s}) - \log (t_1 t_2 t_3) - \log \prod_{i=1}^3 (U_i + \bar{U}_i).
\ee
Relating the matter metrics to the physical Yukawa coupling through (\ref{PhysicalYukawaCouplings}), 
we obtain
\be
\tilde{K}_{12} \tilde{K}_{23} \tilde{K}_{31} \sim \frac{1}{\sqrt{t_1 t_2 t_3}}.
\ee
This is consistent with the exact results \cite{hepth0412150, hepth0404134}, which give
\be
\tilde{K}_{12} \sim \frac{1}{\sqrt{t_3}}, \hbox{ etc}.
\ee

We may also compare with the case of three, differently magnetised, D9 branes wrapping a $T^6$,
a case analysed in full detail in \cite{hepth0404229}.
In this case the chiral fermions have support over the entire $T^6$, with the wavefunctions being
given by
\bea
\psi_{12} & \sim & \frac{1}{\sqrt{V}}, \nonumber\\
\psi_{13} & \sim & \frac{1}{\sqrt{V}}, \\
\psi_{23} & \sim & \frac{1}{\sqrt{V}}.
\eea
Again, in this case we can see from the explicit expressions in \cite{hepth0404229} how the wavefunctions have 
a trivial dependence on the K\"ahler moduli but an intricate (and flavour-sensitive) dependence on the 
complex structure moduli. A sample wave function has the generic form of a product of three functions like (\ref{psisoln}), one for each of the tori in the factorisation\fn{This is in the simplest case of a factorisable $T^6$ with no non-Abelian Wilson lines. As can be checked in \cite{hepth0404229}, the general expression becomes much more complicated.} $T^6=T^2\times T^2\times T^2$. The physical Yukawa couplings thus scale as
\be
\hat{Y}_{\alpha \beta \gamma} \sim \int_V d^6y \, \psi_{ab}(y) \psi_{bc}(y) \psi_{ca}(y) \sim \frac{1}{\sqrt{V}},
\ee
and so the matter metrics behave as
\be
\tilde{K}_{ab} \tilde{K}_{bc} \tilde{K}_{ca} \sim \frac{1}{\sqrt{t_1 t_2 t_3}}.
\ee
These results are consistent with those of \cite{hepth0404229}.

In the toroidal case, the above techniques are not able to fully
determine the matter metrics $\tilde{K}_{ab}$, instead only giving the product
$\tilde{K}_{ab} \tilde{K}_{bc} \tilde{K}_{ca}$. This is a consequence
of the fact that the D7 branes wrap several cycles and that for toroidal examples 
it is not possible to separate the
Yukawa interaction and the overall volume in the same way as for the
large-volume models. It would be interesting if these techniques could be developed to give the
individual matter metrics for the toroidal case.

\section{Discussion and Conclusions}
\label{conclusion}

In this paper we have developed techniques to compute modular
weights  
for bifundamental chiral matter in Calabi-Yau flux compactifications.
These techniques have applications to the computation of soft terms in
gravity-mediated supersymmetry breaking. For chiral matter arising
from a single stack of (magnetised) D7-branes wrapping a small cycle
in the large-volume models of \cite{hepth0502058}, we obtain
\be
\label{ConclResult}
\hat{K}_{\alpha \bar{\beta}} = \frac{\tau_s^{1/3}}{\mc{V}^{2/3}} k_{\alpha
  \bar{\beta}}(\phi) + \ldots,
\ee
where $\phi$ denotes the complex structure moduli. (\ref{ConclResult})
  is the leading term in a series expansion in $\tau_s^{-1}$ and $\mc{V}^{-1}$.

We can see that the bifundamental fields behave like the D3 brane
fields and D7 Wilson lines, in the large volume limit,  rather than
the D7 brane positions. This is consistent
with the fact that bifundamental D7 fields can be localised whereas
the adjoint fields, describing the position of the D7 branes in the
bulk manifold, cannot.

Notice that this behaviour is also different from the one calculated for
toroidal orbifolds. Had we used the toroidal case as
representative of K\"ahler potentials in the general case we
would have been misguided. The reason for the difference is that toroidal
compactifications do not provide good examples of large volume
compactifications where the standard model can be localised on  a D7
brane, independent of the overall volume. The
assumptions we made in the large volume case do not hold for
tori. However we were
still able to use our techniques regarding the structure of Yukawa
couplings  for toroidal cases and get results consistent with the
literature,
 once the peculiarities of toroidal compactifications were
considered.

We have also managed to extend our techniques to give an independent
derivation of the vanishing of the $\mu$ term in the superpotential.
Therefore substantial information can be obtained for Calabi-Yau
compactifications even though explicit string calculations are not
viable. This illustrates the power of our techniques. 

Let us discuss the limitations of the above techniques. First, the 
above method is restricted to modular weights for those moduli that do not appear in the
superpotential Yukawa couplings. Such moduli are those with a shift
symmetry, as these cannot appear perturbatively in the superpotential
and thus in $Y_{\alpha \beta \gamma}$.
The moduli K\"ahler potential is
generally known and this allows the physical Yukawa couplings to be
directly related to the matter K\"ahler metrics. If moduli appear in the superpotential, then it is not possible to
separate the behaviour of the physical Yukawas into superpotential and
K\"ahler potential terms.

However, we still need to know the scaling behaviour of the physical Yukawas.
This gives a second restriction, that the physical Yukawa couplings arise from
essentially classical physics through wavefunction overlap. It is this
that allowed us to compute the scaling behaviour of $\hat{Y}_{\alpha
  \beta \gamma}$ in sections \ref{largevol} and \ref{tori}. If the
Yukawas were nonlocal effects arising from nonperturbative instanton effects 
(as does occur for IIA braneworlds), 
then it is not obvious how to compute the scaling of $\hat{Y}_{\alpha
  \beta \gamma}$. 

Finally, the techniques all apply only to the classical weak-coupling
limit. This is equivalent to determining the leading power $\lambda$ of
$\tau_s$ in the expansion (\ref{MatterExpan}). In IIB
compactifications, $\tau_s$ controls the gauge coupling on D7 branes
and so we expect the full expression of (\ref{MatterExpan}) to be a
series expansion in $\tau_s$.
However if $\tau_s$ ceases to be large, then
the knowledge of simply the leading power $\lambda$ is inadequate as the
expansion is not well controlled. 

We can foresee many applications of our results given the fact that
bifundamental fields are chiral and are expected to provide the
physical observable particles in realistic models. One such
application is to determine the structure of soft terms.
One of the principal difficulties in computing soft terms in the
large-volume models of \cite{hepth0502058} was the lack of knowledge
of the matter metrics for bifundamental fields. This required the
use of generic expressions in \cite{hepth0505076, hepph0512081,
  hepth0605141}, referring to (for example) adjoint matter on D7
branes rather than the bifundamental matter most relevant for the
problem of MSSM soft terms.
This has been addressed in section \ref{largevol} of this
paper 
and the results have obvious applications to the computation of soft
terms that will be presented in a companion paper
\cite{WIP}. 

\section*{Acknowledgements}

We thank S. Abdussalam, C.P. Burgess, L. Ib\'a\~nez,  S. Kachru  and K. Suruliz for
interesting conversations. DC and FQ thank KITP, Santa Barbara and the
organizers of the workshop on `String Phenomenology' for hospitality
during the development of this project. JC is supported by EPSRC and Trinity College, Cambridge. DC is supported by the University of Cambridge. 
FQ is partially supported by PPARC and a Royal Society Wolfson award. This
research was  supported in part by the National Science Foundation
under Grant No. PHY99-07949.

\end{document}